\documentstyle{mn}
\input psfig.sty

\def\h1{$h^{-1}$}

\def\msun{M$_{\odot}$\,}

\title{X-ray properties of the distant cluster Cl0016+16}
\author[Neumann \& B\"ohringer]{Doris M. Neumann and Hans B\"ohringer\\
 Max-Planck Institut f\"ur extraterrestrische Physik,
Giessenbachstr 1.,
85740 Garching}

\begin{document}
\label{firstpage}
\maketitle
\begin{abstract}

We present X-ray data on the distant cluster Cl0016+16 (z=0.5545) from  ROSAT
PSPC and HRI observations and use them to study the physics
of the intracluster medium (ICM) and the dynamical state of the cluster.
The surface brightness distribution is not only described by a spherically
symmetric model but also by a two-dimensional $\beta $-model fit.
Subtracting an elliptical model cluster as defined by the best fit
parameters of the two-dimensional model we find significant residuals,
indicating an additional, extended X-ray source within the cluster.
This source, likely to be a merging subcomponent of the cluster,
coincides with a peak in the weak lensing mass map of Smail et.
al. (1995). In the course of this analysis we present a new approach
to quantify the significance of substructure in cluster X-ray images dominated
by Poisson noise and smoothed with a Gauss filter.\\
We determine the radial mass profile integrated out to a radius of 3Mpc and 
find for the total mass of the cluster a value of  
$\sim 1.4-3.3 \times 10^{15}$ \msun and
$\sim 4.5 \times 10^{14}$ \msun for the gas mass, yielding a
gas--to--total mass ratio of 14 -- 32\%. 
There is no significant radial
dependence of the gas--to--total mass ratio in the cluster.

\end{abstract}

\section{Introduction}

Cl0016+16 is with a redshift of z=0.5545 one of the best studied clusters of 
galaxies at higher
redshifts. It has been target of observations in all wavelength. The cluster
is very massive and, comparing it to nearby clusters, it certainly is
most similar to the Coma cluster. It seems to be embedded in a
large-scale-structure density enhancement
at a redshift of approximately z=0.55 (Koo, 1981). This
idea was recently strengthened by Hughes, Birkinshaw \& Huchra (1995), who
found a poor cluster in X-rays at a redshift of 0.5506 at about 8 arcmin
distance from Cl0016+16.\\
It is an exceptional cluster, due to its high X-ray luminosity
(L$_x$(2-10keV)=2.62$\times 10^{45}$erg/sec (Tsuru et al. 1996) at such a high
redshift. Also the optical appearance of this cluster is different to other
equally distant clusters.
For a long time it was known to be a counter example for the Butcher-Oemler
effect (Butcher\& Oemler 1978) showing only a red, old population of galaxies.
\\
Belloni\& R\"oser (1996) (hereafter BR) recently found that the cluster
has a high fraction of E+A galaxies, being responsible for the relatively high
fraction of red light in the cluster's galaxies. Wirth, Koo \& Kron (1994)
studied 24 cluster members with the HST, and found that the E+A galaxies seem
to be more disklike than normal red galaxies. 
\\
Cl0016+16 is an ideal cluster for the observation of the
Sunyaev-Zel'dovich effect, as it is very
X-ray luminous and at a high redshift. The cluster was among the first three
objects for which a detection of the effect has been claimed
(Birkinshaw, Gull \& Hardebeck 1984; Uson 1986; see also Rephaeli 1995).\\
In this paper we study the X-ray properties of Cl0016+16 using ROSAT/PSPC and
HRI data. We give constraints on the radial total mass profile of the cluster
and discuss its morphology and its dynamical state.
Most probably Cl0016+16 is in the process of a merger. The smaller infalling
component carries only a small fraction of the mass, and
therefore the merger has probably a small effect on the overall dynamical
equilibrium of the cluster.\\
Recently, a weak gravitational shear signal was detected in the cluster
by Smail et al. (1995) (hereafter SEFE). They used this measurement to infer the
gravitational mass of Cl0016+16. We compare our results on the cluster
mass derived from the X-ray data with their weak gravitational lensing 
results.\\
Throughout the paper we use a Hubble constant of H$_0$=50km/sec/Mpc.

\section{The observations}

\subsection{ROSAT HRI data}

\begin{figure}
\psfig{figure=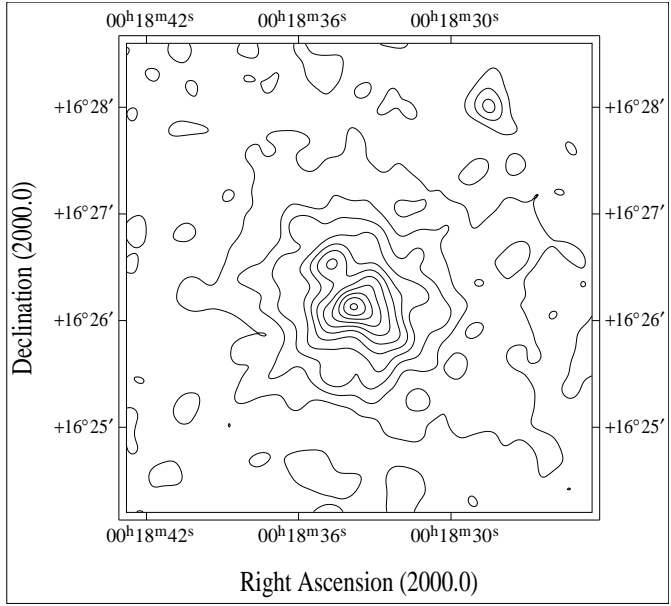,height=8cm}
\caption{Contour Plot of the ROSAT HRI observation of Cl0016+16. The image is
smoothed with a Gauss filter with a $\sigma$ of 6 arcsec.
The lowest contour
level is 1.3$\times 10^{-6}$cts/sec/arcsec$^2$. The spacing is linear with
steps of 5.22$\times 10^{-7}$ cts/sec/arcsec$^2$.}
\label{fig:hriplot}
\end{figure}

Cl0016+16 was observed by the ROSAT/HRI for 76593 sec (for a description of
ROSAT see Tr\"umper 1983, 1992). The observation was
performed in two different periods, in January and June/July 1995 with
exposure times of  5,917 sec and 70,676 sec, respectively.
To correct for a possible offset between the two pointings,
we use the QSO 0015+162 in the North of the cluster.
We measure an offset of $6^{\prime\prime}$ in right ascention and 
$4^{\prime\prime}$
in declination
between both positions of the QSO, and correct for it.\\
Fig.\ref{fig:hriplot}. shows the HRI count rate image of Cl0016+16. The image 
covers  the 
total energy
range of the ROSAT telescope, 0.1-2.4keV.\\
The source in the North-East of the centre is not very significant, and does
not have an optical counterpart (see also Fig.\ref{fig:hriover}.). The 
QSO 0015+162 lies outside of
this plot but can be seen in Fig.\ref{fig:hriplot}.

\subsection{The PSPC data}

\begin{figure}
\psfig{figure=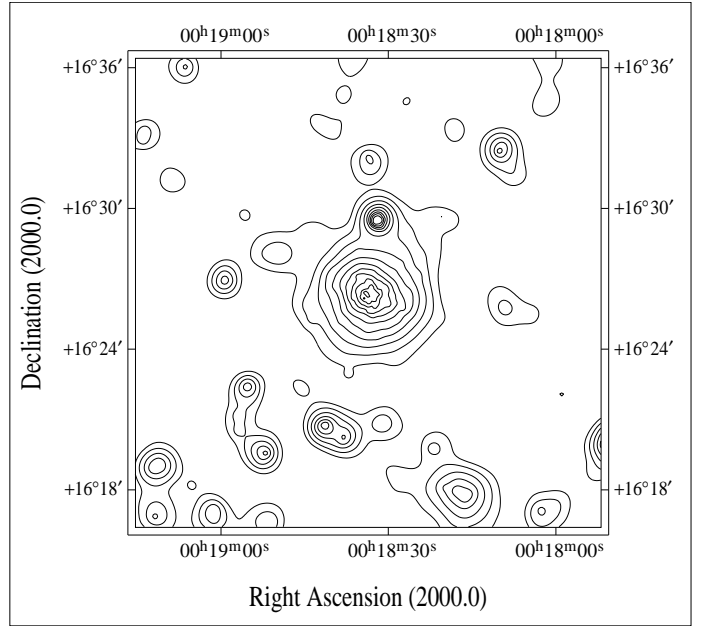,height=8cm}
\caption{Contour plot of the exposure corrected PSPC data of Cl0016+16. The
image is smoothed with a variable Gauss filter. The largest $\sigma$ is 30
arcsec. The lowest contour line is $1\times 10 ^{-7}$cts/sec/arcsec$^2$
The spacing of the contour levels is 35\% of the value of the lower one.}
\label{fig:pspcplot}
\end{figure}

Cl0016+16 was observed for 43,157 sec with the ROSAT PSPC (for more detailed
information see Hughes et al. (1995)). To obtain the best
signal-to-noise-ratio for the detection of the cluster we select the
photons in the 0.5 - 2 keV energy band (channels 52 to 201). The resulting
count rate
image for this energy range is shown in Fig.\ref{fig:pspcplot}.
The image is vignetting corrected. 
The PSPC data show a much more regular appearance, partly coming from the 
larger point spread function (PSF) of the instrument in comparison with the 
HRI, and partly from
the better statistics.
The PSPC has a much lower
background and a higher sensitivity (about a factor of three)
compared to the
HRI.
\\
The most prominent point source in the image is QSO 0015+162,
lying to the North
of the cluster, approximately 3 arcmin  from the centre. In the South,
approximately 8 arcmin  from the centre, one can see a clearly
extended source which is another galaxy cluster at a redshift of z=0.5506
recently discovered by Hughes et al. (1995) in the same PSPC data.
\section{The spatial data analysis}

\subsection{The spherical symmetric fitting}
\begin{figure}
\psfig{figure=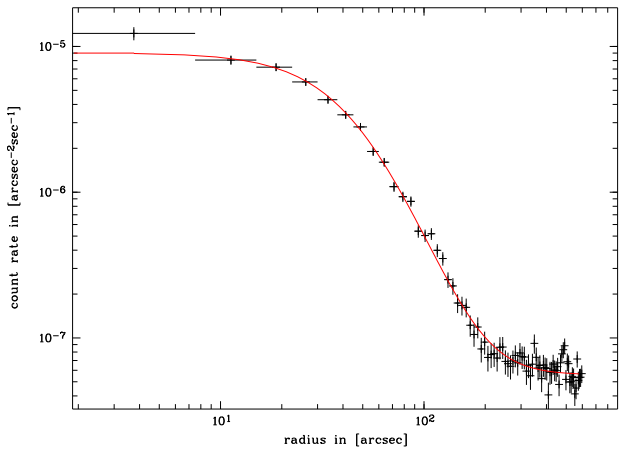,height=6cm}
\caption{Surface brightness profile of Cl0016+16 from the PSPC data. The
crosses mark the errors and the full line is the best fit with the parameters
shown in Tab.\ref{tab:pspc}.}
\label{fig:surf}
\end{figure}

For the data analysis we use EXSAS, the software system provided
by MPE (Zimmermann et al. 1994).
To study the global physical properties of the cluster we primarily use the
PSPC data, because they trace the cluster X-ray emission to larger
radii due to the higher signal to noise and better photon statistics.
But we also present the results of analyzing the HRI data.
With the PSPC data we have also checked for a possible hardness ratio variation 
of the
X-ray emission across the PSPC cluster image and found no variation,
implying that the soft energy band does not provide any particular extra
information on the physical parameters of the cluster.\\
To obtain an approximate description of the X-ray surface brightness
distribution, which allows to subsequently derive the main physical parameters
of the cluster analytically,
we use the so-called isothermal
$\beta$-model (Cavaliere\&Fusco-Femiano, 1976, 1981; Sarazin\&Bahcall, 1977;
Gorenstein et al., 1978; Jones\&Forman, 1984), which describes the surface
brightness $S(r)$ of a galaxy cluster assuming spherical symmetry
\begin{equation}
S(r) = S_0\left( 1 +\frac{r^2}{r_c^2}\right)^{-3\beta+1/2}+B
\end{equation}
where $S_0$ is the central intensity, $r$ the radius, $r_c$ the core
radius, $\beta$ a slope parameter of the radial surface
brightness distribution, and
$B$  the background surface brightness. The $\chi ^2$-fitting
results for the PSPC data are: $\beta = 0.80^{+0.09}_{-0.07}$,
$r_c = 372^{+62}_{-52}$kpc$\equiv 50.5^{+8.4 \prime\prime}_{-7.1}$
(for a summary see Tab.\ref{tab:pspc}). The surface brightness profile and the 
fit are
shown in Fig\ref{fig:surf}. The results for the HRI data are:
$r_c = 283_{-81}^{+140}$ kpc, $\beta = 0.68^{+0.26}_{-0.1}$  (see also
Tab.\ref{tab:hri}). Serendipitous sources are extracted from the fitting.
The best fit values of both data sets
for the core radius and the $\beta$ are
quite different, but the error bars have a
large overlap (see Tab.\ref{tab:pspc} and Tab.\ref{tab:hri}). The discrepancy 
arises from the fact, that the HRI data do not
trace the cluster out to such a large radius as the PSPC data, due to higher
background and lower sensitivity.\\ 
Fitting this spherically symmetric profile to the image of Cl0016+16
has the disadvantage that it does not perfectly describe the slightly
elliptical cluster. But it allows a simple analytic deprojection of
the surface brightness profile of the cluster yielding the radial
gas density profile. We will show later that the spherically symmetric
approximation is a sufficiently good description for the determination
of the gas mass and total mass profiles with errors that are smaller
than relevant uncertainties like those of the temperature distribution
of the intracluster medium (ICM). It can actually be seen also from
the analysis of simulated clusters that a spherically symmetric
$\beta $ model approximation generally leads to a good description of the
shape of realistic clusters with errors of less than 20\% (e.g.
Schindler 1995; Evrard, Metzler \& Navarro 1996). The analytic deprojection
of the profile of equ. (1) leads to the density profile:\\
\begin{equation}
\rho(r) = \rho_0\left(1 + \frac{r^2}{r_c^2} \right)^{-3\beta/2}~~.
\end{equation}
For the central electron density we obtain 0.65$\times10^{-2}$cm$^{-3}$
from the PSPC data, and 0.77$\times10^{-2}$cm$^{-3}$ from the HRI data.\\
For the fitting we cut out regions of serendipitous sources, which clearly do
not belong to the emission of the ICM itself.\\
The different fit parameters are correlated. For example the $\beta$ and the
core radius are dependent on each other. But $\beta$ is also correlated to the
background. A too low
result in the background leads to a decrease of $\beta$, as the profile has to
become shallower to overcome the difference of the wrong background result to
the real one. Therefore we have carefully checked these sources of uncertainty
and found in particular that our result for the background in the fit is in
perfect agreement with the background determined independently in large
regions outside the cluster with all detectable individual sources removed.\\

\subsection{The two-dimensional fit}
To  account better for the slightly elliptical shape of the cluster we have
also performed a two-dimensional fit to the cluster using a modified
$\beta $ model that allows for two different core radii along
the two principal axes of the cluster image ellipse. The surface
brightness profile in this model is described by the equation:\\

\begin{equation}
S(r_1,r_2) = S_0 \left(1 + F_1 +
F_2\right)^{-3\beta+1/2}+B
\end{equation}
with
\begin{displaymath}
F_1 = \frac{(\mbox{cos}(\alpha)(r_1-x_0)-
\mbox{sin}(\alpha)(r_2-y_0))^2}{a_1^2}
\end{displaymath}
\begin{displaymath}
F_2 = \frac{(\mbox{sin}(\alpha)(r_1-x_0)+\mbox{cos}(\alpha)(r_2-y_0))^2}{a_2^2}
\end{displaymath}
The fit is applied to the two-dimensional  pixel data of the image.
The newly introduced parameters are defined in Tab.\ref{tab:defi}.

\begin{table}[h]
\caption{Description of the fit parameters of the two-dimensional case. Not
mentioned parameters have the same definition as in the 1d fit.}
\begin{tabular}{l|c}
parameter & definition \\
\hline
$x_0$       & central position in x direction \\
$y_0$       & central position in y direction \\
$r_1$       & coordinate in  x direction \\
$r_2$       & coordinate in y direction   \\
$\alpha$  & position angle \\
$a_1$     & major axis of core radius \\
$a_2$     & minor axis of core radius  \\
\end{tabular}
\label{tab:defi}
\end{table}

The results are shown in Tab.\ref{tab:pspc} and Tab.\ref{tab:hri}.
This fit has the advantage of taking into account, that a relaxed cluster
does not necessarily need to be completely spherical symmetric, but can show
ellipticity. A further advantage is that the centre position
of the cluster is derived in the fitting procedure itself and does not have
to be predetermined like it is necessary in the
one-dimensional analysis.
There the centre position is usually determined from the maximum in
the surface brightness profile (with some dependence on the smoothing used
before determining the centre).\\
The application of such a two-dimensional fit can also be found in
Bardelli et al. (1996) in which they apply this model
to Abell 3558 the central
cluster of the Shapley concentration.\\
However, the two-dimensional fit exhibits more problems than the
one-dimensional one, firstly because of poorer statistics in each bin,
as one has to use pixels instead of concentric rings. This is particularly
severe at larger radii because the surface brightness is falling off steeply,
which is partly compensated by the increasing surface of the rings in the
one-dimensional analysis but not in the two-dimensional case.
This is very crucial in the regions outside the cluster
emission, where the only observed emission comes from the background.
Secondly, the number of fit parameters increases from four to eight fit
parameters.\\
For our two-dimensional fit we use the PSPC data with a binsize
of $5^{\prime\prime}\times5^{\prime\prime}$. We again cut out serendipitous
sources in the
field of view. Since the number of photons per pixel in the background area
is too small to apply Gaussian statistics, which is assumed by
$\chi^2$-fitting, we apply a small Gauss-filter to the image.
The $\sigma$ of the Gauss-filter is $5^{\prime\prime}$, and therefore much smaller
than the
FWHM of the PSF ($\sim 30^{\prime\prime}$) of this data set, so that we do not
risk to
loose any information.\\
If we do not apply a Gauss-filter, we underestimate the background.
This is due to the fact that for the low photon statistics in the
background pixels the mean of the Poissonian distribution is larger
than the most abundant value and the mean adopted by assuming a Gaussian
distribution. This adopted Gaussian distribution is used with $\chi^2$-fitting.
Because of the above mentioned correlation effect
an underestimation of the background leads than to a decrease in the
fitted slope parameter $\beta $. The application of a small Gauss
filter improves the photons statistics and removes this problem.
Of course, the errors obtained for the fitting parameters in this
procedure are no longer precisely defined.  However, it is a very
good approach to get the best fit value. We test this with images on model
clusters, on which we added Poisson noise. These model clusters have about
the same parameters as Cl0016+16. We apply a Gauss-filter, of the same size as
for the real cluster image, and run the fitting routine on them. The parameters
obtained from the fitting are in very good agreement with the intrinsic
model parameters, so that we can be sure, that the Gauss-filter does not
obscure our fitting result.\\
The one-dimensional fit-parameters are in very good agreement with the
two-dimensional fit-values (see Tab.\ref{tab:pspc} and Fig.\ref{fig:masse}), so that this can also be
used as a proof for reliability of the Gauss filtering. The best fit
value for the core radius in the one-dimensional fit is 372 kpc, lying almost
exactly on the geometrical mean value of the 2d fit. Also the $\beta$ of the
two fits agree very well, with 0.80 best fit value for the 1d fit, and the
2d fit being at 0.81.\\
Trying to overcome the Gauss filter with a larger binsize of the image is not
feasible, due to the low background. The bins would be too large.
The best fit value for from the two-dimensional fit are slightly
depending on the binning. However, applying different binning and different
Gauss filters (not too large of course) yields values which are all in
excellent agreement with the 1d fit and its errors.\\
Another approach, using a maximum likelihood analysis, as for example applied
by Birkinshaw, Hughes \& Arnaud (1991), has the disadvantage of predefining the
fit parameters. The parameters are not fitted, but only tested. \\
To check the existence of substructure, in addition to the overall ellipticity 
of the
cluster, we produce a synthetic cluster image from the fit parameters of the
two-dimensional model and subtract it from the real image. There is
significant structure in the residual image that can indicate
either point sources in the field or subcomponents of the cluster
itself. The residual images are shown in Fig.\ref{fig:pspcover} for the PSPC 
data and
in Fig.\ref{fig:hriover} for the HRI data. To obtain Fig.\ref{fig:hriover}
the HRI image was 
treated
in the same way as the PSPC image with a two-dimensional fitting approach.
The used parameters are shown in Tab.\ref{tab:hri}. Here the pixel size is 
$4^{\prime\prime}\times4^{\prime\prime}$ and also $\sigma=4^{\prime\prime}$
for the Gauss filter.
The residual images shown in these figures are displayed in the form of
a significance plot. How the significance of the pixels of the residual
images are calculated is explained in more detail in the appendix. Briefly
the initial problem is how to calculate the significance of a source after
having applied a Gauss filter. We solve this problem by using error propagation
and Poisson noise. Basically two different Gauss filtered images are divided
by each other. The one being the ``error image", is not only Gauss filtered,
but also the square root of each pixel has to be calculated together with
a general normalization. The ``source image" can  be calculated in two
ways. In the first case one subtracts a background model. Then the routine gives
$\sigma$ above background. The other possibility is to subtract a model cluster,
and to calculate the significance ($\sigma$) of these residuals.
A more general approach for the testing of significances with filter techniques
(not only Gauss filters) is described by Rosati (1995).

\section{The spectral analysis}

Due to the spectral resolution of the PSPC it is possible to determine the
temperature of the ICM. However, as this cluster has a very high redshift
and only a relatively small number of source photons were detected, the
spectral resolution is of course limited.\\
We determine the temperature in different radii around the centre of Cl0016+16
with different backgrounds and different energy bands with different values for
metallicity. The results show a
large scatter, which is due to the relatively low photon statistics. For
example in a radius of 4 arcmin around the centre, the total sum of
photons is about 6,000.\\
Nevertheless, the results which are in the range 6 to 10 keV are
in quite good
agreement with the ASCA data of Tsuru et al. (1996) who obtained a global
temperature of $8.22^{+1.83}_{-1.47}$ keV. Only the results for the hydrogen
column density are different. We obtain results of $4-6\times10^{20}$cm$^{-2}$,
which are in good agreement with the measurements of the 21cm line of
$4\times10^{20}$cm$^{-2}$(Dickey
\& Lockman 1990), while Tsuru et al. find  $12.9\times 10^{20}$cm$^{-2}$.
This difference might be explained by the fact,
that ASCA does not provide a very good spectral sensitivity below 0.5keV, which
is the energy range relevant for the determination of this degree
of absorption.\\
In different rings (which are partly overlapping) we do not detect any
significant temperature gradient,
which leads us to the assumption that the cluster is probably more or less
isothermal within the given the large error bars. This is 
in very good agreement with
the ASCA data. Only in the very centre we see a drop of the lower boundary of
the temperature (in a radius of 1 arcmin) to 4 keV together with an
increase of the hydrogen column density. This decrease is, however, not 
significant and
most likely an artifact of the temperature fitting,
as the cluster has not yet developed a Cooling Flow (see also 5.2)
for which such a temperature decrease would be expected. Therefore
we neglect this lower boundary in the centre for the mass determination.
Including it does not change the overall mass result significantly (the mean
result changes by less than 5\%). \\
In general it is interesting, that the temperatures determined from the
PSPC data are only very weakly depending on the assumed values for the
metallicity.

\section{The Morphology of Cl0016+16}
\subsection{The X-ray data}
Cl0016+16 shows a position angle of about 50$^{\circ}$
(measured from North to East) and
an eccentricity of about 0.18. There are clearly additional
sources superposed on the elliptical cluster, both
   in the HRI and the PSPC observation.\\
\begin{figure}
\psfig{figure=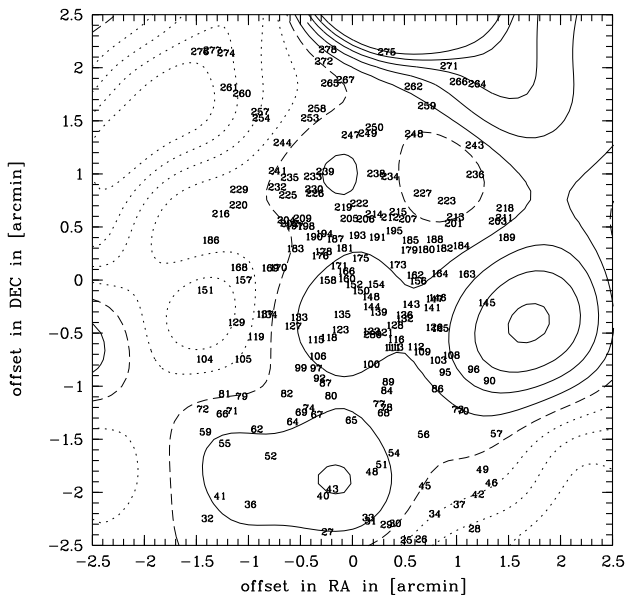,height=8cm}
\caption{Significance of the residuals after subtracting the elliptical
isothermal $\beta$-model from the original hard PSPC image. The parameters for
the subtracted model is described in Tab.\ref{tab:pspc}. The stepsize of the 
contours
is 1
$\sigma$. The dashed line is the 0 level, the full lines are positive the
positive $\sigma$'s, the dotted ones are the negative ones. The numbers are the
objects of the optical identification list of BR. The Gauss
filter has a $\sigma$ of 25 arcsec.}
\label{fig:pspcover}
\end{figure}
\begin{figure}
\psfig{figure=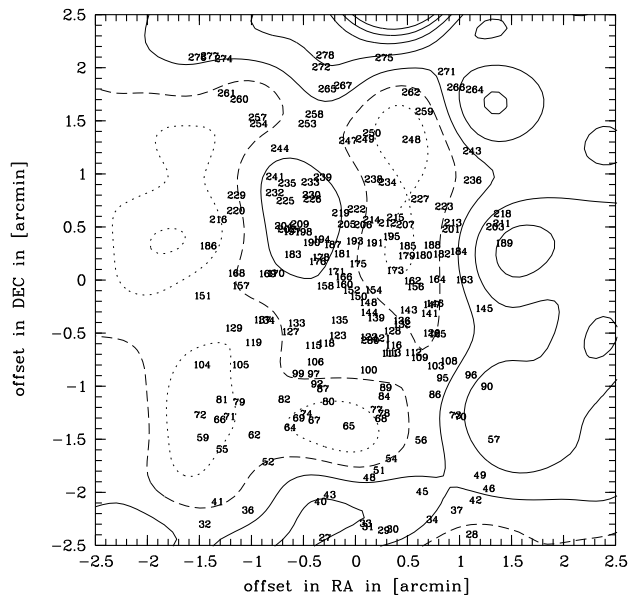,height=8cm}
\caption{Significance of the residuals after subtracting the elliptical
isothermal $\beta$-model from the original HRI image. The parameters are
presented in Tab.\ref{tab:hri}.
The stepsize is 1
$\sigma$. The dashed line is the 0 level, the full lines are positive the
positive $\sigma$'s, the dotted ones are the negative ones. The numbers are the
objects of the optical identification list of BR. The Gauss
filter has a $\sigma$ of 20 arcsec.}
\label{fig:hriover}
\end{figure}
In the West of the cluster, about 2 arcmin  from the centre
(Fig.\ref{fig:pspcover}), we find
a surface brightness excess with 4 $\sigma$ significance in the PSPC image.
This source at a position of RA=00H18M27S DEC=16D25M55S (J2000)
is most likely extended, as explained in the discussion.\\
The residual image of the HRI observation also shows a surface brightness
excess to the West of the centre. At the exact location of the PSPC maximum
of Fig.\ref{fig:pspcover} we find, however a gap in the HRI surface brightness 
excess (Fig.\ref{fig:hriover}).
There are several effects that can produce such a result. First
of all, the HRI data have a lower photon statistics than the PSPC data
as explained above and thus the gap may be produced by a
larger statistical fluctuation
(the surface brightness excess is just produced by 55 photons).
The second point is, that only the hard band photons were used for
the analysis of the PSPC data.
We cannot apply the same selection to the
HRI, as this instrument does not provide any energy discrimination.
But as we found no significant feature in a hardness ratio map produced
from the PSPC data we believe that the difference in the shape of the
western emission excess is due to photon statistical effects.
\\
Therefore we conclude, that there is an extended source in the
West in the
line-of-sight of Cl0016+16. This extended source might be a subgroup falling
into this cluster. However, this subgroup, if existing is very small compared
to the total cluster Cl0016+16.\\
Another source visible in the significance plot of the PSPC image, but only
having a 2$\sigma$ significance, is the
source in the South. This source structure is most likely to be extended too.
Galaxy number 43, defined by BR, having a redshift of z=0.38, indicated
in the figure, coincides
with the maximum of this source, so that it might be, that there is emission
from a small foreground group, as already suggested by Ellis et al. (1985).
However, optical source number 36 also in this
enhancement is a star and might contribute partly to the emission.

\subsection{Cooling time}

The central cooling time, calculated for a gas temperature of 8.4keV
(Tsuru et al, 1996; Yamashita, 1995) and for
the density profile obtained from the isothermal $\beta$ model fit
is larger than 10$^{10}$years from both the PSPC and the HRI data. Therefore 
the cluster cannot have yet formed a steady cooling flow (Fabian, Nulsen \&
Canizares 1984). 

\subsection{Comparison with optical data}

Cl0016+16 is a well observed cluster in the optical. There has been an
extensive survey by Dressler \& Gunn (1992) (hereafter DG) and BR.
Fig.\ref{fig:pspcover} and Fig.\ref{fig:hriover} show the overlays of the 
optical data of the sample of
BR.\\
We did not apply a bore-side correction in the PSPC image, but found an
inaccuracy of the data of 2 to 5 arcsec for the pointing position.\\
The most striking feature is, that the alignment of the central galaxies is 
coinciding with the PA we determined from X-rays (as already mentioned by
SEFE).  However, for the
4 $\sigma$ excess emission in the West of the centre, which is
probably a substructure
feature of the cluster, there is not enough overlap with the data of BR
to search for a possible galaxy density enhancement which could be
related to the surface brightness excess.
DG detected objects in that region, but they did not obtain
redshifts for them. The objects seem to be extended, so that it is likely that
they are galaxies.\\

\section{Mass analysis}

\subsection{The one-dimensional approach}

X-ray observations of the ICM in clusters of galaxies can be used  for the
determination of the total and gas mass.
The formula for the cluster mass profile obtained from combining
the isothermal $\beta$ formalism (1d) equ(2) and the hydrostatic equation is:
\begin{equation}
M(<r) = \frac{k r^2}{G \mu m} \left(\frac{dT}{dr} - \frac{3 \beta r
T}{r^2+r_c^2}\right)
\end{equation}
The results are shown in Fig.\ref{fig:masse}. For the mass determination we use 
the
PSPC data because of the better statistics, even though the
spatial resolution is a factor about 7 worse in comparison to the HRI data. Our
total mass estimate for Cl0016+16 at a radius of 3Mpc is about 2.3$\times
10^{15}$
\msun. This is a typical result for a rich cluster, like for example the Coma
cluster of galaxies (Briel, Henry \& B\"ohringer 1992). \\
For the mass determination we use the method by Neumann \& B\"ohringer (1995).
This method uses a Monte-Carlo approach for determining the cluster's radial
mass profile. The stepwidth for the Monte-Carlo method is 300kpc with a
maximal change of the temperature of 1 keV per step. For the
temperature range we take 6 to 10 keV (rounded errors of Tsuru et al. 1996 and
our results from the PSPC data fitting). We assume that the cluster gas is
not necessarily isothermal, but only that the gas temperature
lies in the given temperature range.
This assumption is justified since we find no significant temperature
variations in different concentric rings.
The gas mass of the cluster integrated out to a radius of 3 Mpc is roughly
4.5$\times10^{14}$ \msun . Thus we obtain a  gas to total mass ratio of about
20\%. This is in the typical range of other clusters.

\subsection{The two-dimensional approach}

With our results for the two-dimensional fit, it is also possible to determine
the influence of ellipticity on the results for the cluster mass.
For comparison we determine the masses with the major and minor-axis
results for the core radius of the two-dimensional
fit. We again use the Monte-Carlo approach with the same conditions as the
one-dimensional
analysis. As one can see, the difference between both results becomes less with
increasing radius. This result is natural, as only the coreradius is different,
which influences the mass result mostly near the centre.

\subsection{Comparison of the different mass estimates}

\begin{figure}
\psfig{figure=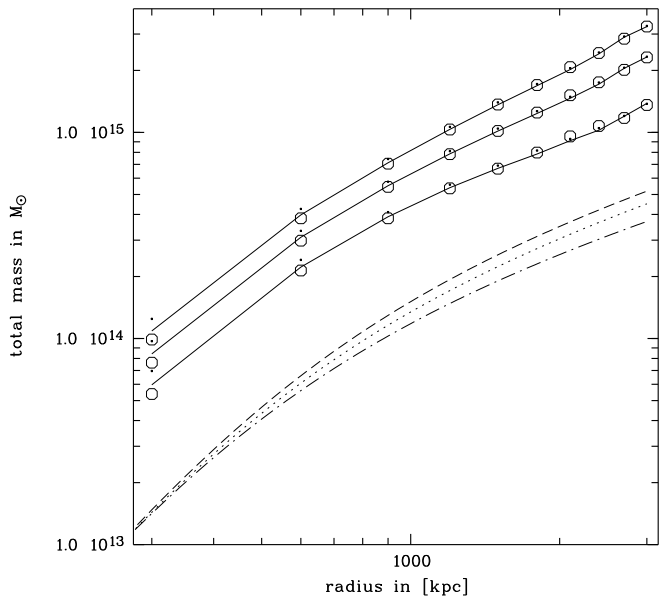,height=8cm}
\caption{Integrated mass profile of CL0016+16. The full lines indicate the
mass profile from the 1d fit. It shows the mean value and the upper and lower
lines are the mean $\pm 2 \sigma$. The small dots are the results for the 2d
fit with the minor axis as core radius. The circles are the same for the
major axis as core radius. Again middle values are the mean, the others $\pm 2
\sigma$. The lower dotted line is the gas mass profile from the 1d fit, the
dashed line is the gas mass from the 2d fit, using the major axis value as core
radius, the dotted dashed line respectively for the minor axis 
result used as core
radius.}
\label{fig:masse}
\end{figure}

Comparing the different results for the total mass of the 1d approach with the
2d approach leads to similar results, as can be seen in Fig.\ref{fig:masse}.
 This also implies that
it is a good approach even to take only a sector out of a certain cluster
surface brightness distribution when the cluster is elliptical. This approach
has been normally undertaken, if clusters show clear substructure, to 
exclude those
regions. Only within the innermost 1 Mpc there is some divergence of the
mass depending on the core radius. However averaging the 2d distribution to a
modified 1d distribution leads to a core radius (by taking the geometrical mean
of the major and minor axis) of less than 1\% difference to the coreradius
originally
derived by the 1d fit. This is a proof, that the 1d fit averages correctly
over the eccentricity of a cluster. Also the similarity of the $\beta$'s of
both approaches supports this.\\
A comparison of the results for the gas mass shows that this case is slightly
more critical.
Because of the dependence of the central density on $\beta$ and the
coreradius, one yields different results for the approach with the values for
the
major and minor axis. Therefore an analysis, that takes into account only 
sectors of the
X-ray distribution must be undertaken cautiously, not to overestimate or
underestimate the mean coreradius. Taking the extreme values for
the major and minor axis, we get a difference to the averaged value almost up
to 20\%.\\
However, averaging over the {\it entire} cluster with a 1d fit is a good 
approach as the {\it averaged} 2d central
electron density is of the order of 1\% different to the one from the original
1d approach. Therefore the
differences of the averaged 1d approach to the original 1d fit are in the order
of less than 1\%, and only diverge to less than 1.5\% at a radius larger than
2.5 Mpc.\\
\begin{table}
\caption{Obtained parameters for Cl0016+16 using the PSPC data. The physical
values are all calculated with $H_0$ = 50 km/sec/Mpc. The errors for $r_c$ and
$\beta$ from the 1d fit are 3$\sigma$ errors.}
\begin{tabular}{lc}
parameter & value \\
\hline
$\beta$ 1d fit &   $0.80^{+0.09}_{-0.07}$      \\
$r_c$   1d fit &   372$_{-52}^{+62}$ kpc $\equiv
50.5_{-7.1}^{+8.4\prime\prime}$
 \\
$n_{e0}$ 1d fit &  0.65 $\times 10^{-2}$cm$^{-3}$      \\
$M_{tot} <$ 3 Mpc & $2.32^{+0.94}_{-0.94}\times 10^{15}$ \msun    \\
$M_{gas} <$ 3 Mpc &  $4.5\times 10^{14}$ \msun \\
\hline
$\beta$ 2d fit & 0.81            \\
$r_c$ for major axis &  410 kpc $\equiv 56^{\prime \prime} $ \\
$r_c$ for minor axis &  337 kpc $\equiv 46^{\prime \prime} $ \\
PA (North over East) &  50$^{\circ}$ \\
$\epsilon$ (eccentricity)         & 0.18 \\
\end{tabular}
\label{tab:pspc}
\end{table}
\begin{table}
\caption{Obtained parameters for Cl0016+16 using the HRI data. The errors for
$r_c$ and $\beta$ from the 1d fit are 3$\sigma$ errors.}
\begin{tabular}{lc}
parameter & value \\
\hline
$\beta$ 1d fit &   $0.68^{+0.26}_{-0.1}$      \\
$r_c$   1d fit &   283$_{-81}^{+140}$ kpc $\equiv 
38.5_{-11}^{+19 \prime\prime}$
   \\
$n_{e0}$ 1d fit &  0.76 $\times 10^{-2}$cm$^{-3}$      \\
\hline
$\beta$ 2d fit & 0.69            \\
$r_c$ for major axis &  318 kpc $\equiv 43^{\prime\prime}$  \\
$r_c$ for minor axis &  261 kpc $\equiv 35^{\prime\prime}$  \\
PA (North over East) &  50$^{\circ}$ \\
$\epsilon$ (eccentricity)         & 0.18 \\
\end{tabular}
\label{tab:hri}
\end{table}
\section{Discussion}

\subsection{Possible influence of the QSO on the X-ray temperature}

Tsuru et al. (1996) found with ASCA a temperature of about 8.22keV 
and a very low metallicity, m$<$0.167 for the ICM.
However, these values can in principal be obscured as there is another X-ray 
emitting 
source, QSO 0015+162 in the vicinity of the cluster (see also
Fig.\ref{fig:pspcplot}).
The distance  to the cluster centre of
3 arcmin. This is in the order of the half power diameter of the point
spread function of the ASCA plus
GIS combination of 3.2 arcmin (Ikebe, 1996). Tsuru et al. 1996 used the GIS
data. It is
very likely, that this QSO, which has a redshift of z=0.553, close to the
cluster's redshift, influences the spectrum and therefore the
spectroscopic estimates
might be wrong. A rough estimate of the influence can be made by comparing the
countrate of the QSO with the countrate of the cluster itself measured by
ROSAT.
For the PSPC
we get a countrate of about 10\% for the QSO in comparison to the cluster.
Fitting a power law to the QSO
spectrum of the PSPC yields a power law index in the range of -3 to -2.3.
Fitting a power law to the cluster itself yields a power law index of -1.6 to
-1.5, both times fixing the hydrogen column density to 5$\times
10^{20}$cm$^{-2}$. Thus the spectrum of the QSO is steeper implying that it has
less
influence in the higher energy band of ASCA.
For the
temperature measurement, the QSO would, if it influences the cluster spectrum,
lower the fitted temperature, as fitting
a Raymond-Smith spectrum to the QSO data yields a best fit parameter of about 2
keV.
\\
Another possibility
in principle is
to measure the influence of the QSO to the cluster spectrum by comparing
luminosities measured with ASCA and the PSPC. Tsuru et al. (1996) obtain a
value for $L_x(2-10$keV$)= 2.62\times 10^{45}$erg/sec. With the PSPC we get
$L_x(2-10$keV$)= 2.3-4.4\times 10^{45}$erg/sec. This shows that the errors are
too large to give a definite result on the influence of the QSO. For the 
luminosity determination we took
into account variations of the hydrogen column density, the metallicity and the
temperature (6--10keV).\\
Taking this all together, it is not likely, that the QSO contributes
sufficiently
to the spectrum of the cluster, to explain the very low result for the
metallicity. A 10\%
contribution to the continuum of the ICM spectrum probably does not lower the
metallicity by a factor of two to three, which would be necessary, to make
Cl0016+16 a `normal' distant cluster, with a metallicity of about m=0.3. \\
The low value for the metallicity is rather surprising, as Cl0016+16 is the 
only cluster showing such a result. This does not fit in the normal scenario 
of metal enrichment,
especially as the cluster exhibits many E+A galaxies (BR), which have already
burned all O and B stars. These stars would have been
in principle able to enrich the ICM by their supernovae. However, this does
not seem to be the case.

\subsection{The dynamical state of Cl0016+16}

The X-ray image provides strong evidence for the existence of substructure in
the cluster which is possible due to the merging of a galaxy group with the
main cluster.
\\
In this case, there are basically two possibilities to explain the
observation, as suggested by simulations (Schindler \& M\"uller 1993). The first
one is, that the subgroup is on its first infall to the
cluster, the second one is, that the subgroup already penetrated the cluster
centre once.\\
The second scenario leads to a much stronger distorted appearance of the
subgroup than the
first one. As the subgroup can be seen as a more or less compact feature, the
first case seems to be far more likely.
Another very interesting feature is the chain of galaxies, which coincides
very
well with the PA of the gas. It is not clear whether this chain comes from a
recent merger, and if, whether the measured substructure in the West has
something to do with it. The merger direction of the substructure roughly
agrees
with the PA of the gas. This is a common feature for mergers as predicted by
simulations (Van Haarlem \& Van de Weygaert, 1993). However, it might also be,
that the chain of galaxies are only
the remnants of a former merger, proving that the cluster formed out of
filaments.\\

\subsubsection{Mass estimate of the possible subgroup}

The subgroup in the West shows up with about 55 photons in the hard ROSAT/PSPC 
band of 0.52--2.01 keV.\\
Due to the limited number of photons, it is very difficult to give estimates
on the physical quantities of this source. However, assuming this source is at
the cluster's redshift, and on its first infall to the cluster, so that it
still contains most of its own X-ray emitting ICM, we obtain a
X-ray luminosity of about $L_x\sim
10^{43}$erg/sec. Comparing this luminosity to groups of galaxies, we find,
that this value is on the upper limit for Hickson groups (Ponman et al. 1996).
Poor clusters of galaxies, having more galaxies than Hickson groups are lying 
in the regime of about
10$^{44}$erg/sec, depending if they are hosts for Cooling Flows or not.
Therefore we can classify our source as a group of galaxies in the range
between Hickson groups and poor clusters of galaxies. The mass range, which one
observes for these objects lies in the order of 10$^{13}$ \msun  to 10$^{14}$ 
\msun (Ponman et al. 1996; Mulchaey et al. 1996). This is the 
most likely mass range for the
subgroup. Comparing this result with the total cluster mass of about 2$\times
10^{15}$ \msun, we find a
contribution of the subgroup to the total mass of about 1\% to 20\%.

\subsubsection{Contamination by stars}

\begin{figure}
\psfig{figure=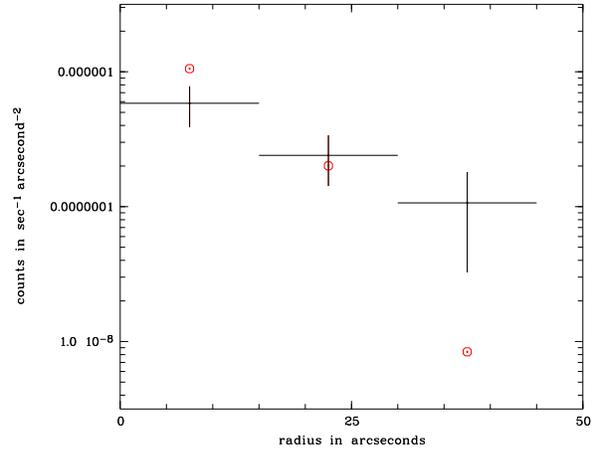,height=6cm}
\caption{The surface brightness profile of the additional source in the West
of the centre of Cl0016+16. The crosses mark the data, inclusive errors, the
dots show the distribution of a point source having the same number of photons
as the original residual source.}
\label{fig:point}
\end{figure}

It must be noted, however, that this additional source does not need to be a
cluster group at the same redshift as Cl0016+16. It might be a chance alignment
 of two point
sources, or it might be a group of galaxies at another redshift. A single
point-source is pretty unlikely to be the case, as we can reject a point-like
shape with a confidence limit of 99\%. This confidence limit is determined by
comparing the surface brightness distribution of the source with the surface
brightness distribution of an artificial point-source having 55 photons
(see also Fig.\ref{fig:point}).
The surface brightness distribution is marginally affected by the (subtracted)
model parameters and the pixel size of the image.
\\
To exclude the possibility of an artifact produced by the satellite observatory
(e.g.
wobbling),  we compare the quasar in the North of the cluster with a point
source, again using the surface brightness distribution. We obtain very good
agreement of the quasar with a point-source, again
the artificial point source having the same number of photons than the quasar
itself, after subtracting the cluster emission (note: the Poisson errors are
calculated for the total number of photons per bin, not only for the residual
photons).
Inspecting the optical image we find at a distance
of about half an arcmin a red star. This star is by far the brightest optical
source
in this region. It is a M4V dwarf (BR).
We can calculate an upper
limit for the X-ray luminosity of the star, using the magnitude of
the luminosity in the red band of m$_r$=16.84, and m$_v$=18.5 in the visible
(from the Cosmos data base).
The saturation limit for M dwarfs is found 
to have a 
value of
10$^{-3}$ (Fleming et al. 1993). This value defines the upper limit of
the energy radiated in
X-rays versus the energy emitted in the optical for M dwarfs.
The 55 excess photons found in the hard
ROSAT band exceed this limit by
a factor of 8, excluding this star as the source of the X-ray emission.\\
However, we cannot exclude Quasars or AGN's to be the
source, probably, if the case,  in a chance alignment.\\
A strong indication for the existence of a group of galaxies in the West of the
centre of the cluster can be found by comparison with the
dark matter map obtained by SEFE using the weak lensing approach
by Kaiser \& Squires (1993).

\subsection{Comparison with weak lensing}

Recently SEFE presented a map of the dark
matter distribution
measured by weak lensing. They also made a rough estimation of the mass
inferred by X-rays, also using the ROSAT/PSPC data.\\
First of all, as SEFE already discussed, the PA of the mass distribution the 
light
and the X-ray emission agrees well. 
Comparing our mass results with the SEFE
mass estimates, however, yields a discrepancy. Our result of
the {\it projected} mass is 4--7$\times10^{14}$\msun integrating out to a 
projected
radius of 600 kpc, whereas SEFE derive a
projected mass of 7.3$\times 10^{14}$\msun using X-rays and a result for the
weak lensing mass of 8.5$\times 10^{14}$ \msun (depending on the
mean redshift of the lensed background galaxies) out to the same distance.
This discrepancy can be
explained by using a different projection model for the X-ray surface
brightness, which is more peaked in the centre than observed with the limited
resolution
of the PSPC data. Also shifting the lensed background galaxies to redshifts at
1$<$z$<$1.4 can in principle resolve the discrepancies. The fact that the
background galaxies might be at a higher redshift than 1 is also supported by
an observation of Luppino \& Kaiser (1996). They measured the weak shear around
MS1054-03, a rich and very X-ray luminous cluster at z=0.83 (Luppino \& Gioia
1995), and found a 
mass-to-light ratio in 
agreement with more nearby clusters, when the lensed background galaxies lie
at a redshift of about 1.5.
\\
Comparing the morphology of the weak lensing mass map with our
residual images yields very good overall similarities. The bimodality which 
can be seen in the weak
lensing maps exhibits similarities as seen in Fig.\ref{fig:pspcover}, and also 
less striking in Fig.\ref{fig:hriover}. The structure in the West
found in the PSPC data coincides well with the western extension in the map of 
SEFE.

\subsection{The gas to total mass ratio in Cl0016+16}

Integrating over the entire cluster (outer boundary 3Mpc), the gas to total
mass
ratio of the cluster is 14--32\%. This is calculated for the 1d fit.
We did not take any uncertainties concerning the
hydrogen column density into account. For the overall temperature we took 8.4
keV. These results are typical for clusters and undeline the effect of the
Baryonic Catastrophe (Briel et al. 1992; White et al. 1993). The
ratio does not vary very much with radius, it stays constant within the error
bars.

\subsection{The Eccentricity of Cl0016+16 and its consequences}

In X-rays (ROSAT/PSPC and HRI data) Cl0016+16 shows an eccentricity of
$\epsilon$ = 0.18. As HRI and PSPC show the same eccentricity, the larger PSF
of the PSPC does not seem to affect this value.
It is in good agreement with the optical distribution of the galaxies of
$\epsilon_{gal}=0.21\pm0.02$ derived by SEFE. The eccentricity is also
comparable
with results for other clusters like A2199 (Gerbal et al. 1992) and the sample
of Buote \& Canizares (1992) (hereafter BC). BC's results on $\epsilon$ are
generally smaller than the value for Cl0016+16, but this might be caused by the 
limited area they took into account for the determination (in all cases 
$<$ 1Mpc).\\
As we proof, the ellipticity of Cl0016+16 hardly affects the total mass
profile of the cluster. The fact that the galaxy distribution and the X-ray
surface brightness lead to similar eccentricities is a strong argument that
we do not underestimate the eccentricity of the dark matter, despite the fact
that SEFE obtain a result for the eccentricity of the dark matter of
$\epsilon_{wl}=0.61\pm0.06$ ($r<<r_c$) via the weak
 lensing approach of Kaiser \& Squires (1993). This discrepancy might be
caused by the limited area of the CCD's for the lensing approach. Therefore it
might be, that SEFE only measure the eccentricity in the centre, which is
likely to be higher than the rest of the cluster, due to the chain of galaxies,
and the results of BC, who found that clusters in X-rays tend to show a  
rounder
appearance with increasing radius. It is also possible, that the eccentricity
measured by SEFE is partly overestimated due to the clear bimodal shape of the 
dark mass map.
Generally
BC found out that X-rays are a good tracer for the shape of the dark matter,
better than for example the galaxy distribution.
However, the influences on the gas mass profile are
relatively high (deviations of up to 20\%). Thus taking for the gas mass
estimate only sectors of the whole cluster can bias the results. This is an
approach X-ray astronomers sometimes undertake, to eliminate regions which are
strongly affected by substructure.
However the
differences are not drastic enough to change the gas-to-total-mass of the
clusters dramatically.

\section {Summary}

In this paper we present X-ray properties on Cl0016+16, a distant cluster of
galaxies with a high X-ray luminosity of 
L$_X$(2--10keV)=2.3--4.4$\times10^{45}$erg/sec. 
The cluster has a temperature between 6-10keV. ASCA and ROSAT
temperature results being well consistent. 
We present the total mass 
profile of
the cluster based on X-ray measurements on ASCA and ROSAT, using a Monte-Carlo
technique.
The mass lies
between 1.4--3.3 $\times 10^{15}$ \msun integrated out to a radius of 3Mpc.
The gas-to-total-mass ratio lies between 14-32\%, making this cluster another
example for the so-called Baryonic Catastrophe (Briel et al. 1992; 
White et al. 1993).\\
Applying a substructure analysis, we find indications for a not very massive 
group of galaxies falling into the centre of the cluster. This analysis is
based on subtracting an elliptical model cluster
following the isothermal $\beta$-model from the cluster X-ray image,
with subsequent testing of the residuals.
 This indicates, that Cl0016+16 is not very disturbed.
The PA's of the distribution of the E and E+A galaxies, the weak lensing
mass distribution, and the X-rays are coinciding very well with about 
50$^\circ$.

\section*{ACKNOWLEDGMENTS}

We like to the thank you the ROSAT team for providing us with the necessary
tools for reducing the data.  We like to thank
Sabine Schindler and Rien van de Weygaert for useful discussions.
We are very grateful to Paola Belloni, for very helpful discussions and
providing us with the positions of the optical data. We also like to thank
Chris A. Collins and Thomas Bergh\"ofer for useful comments.
DMN is very grateful for the hospitality of
the Kapteyn Institute in Groningen.\\
HB is supported by the Verbundforschung of DARA.
ROSAT is supported by BMFT.\\
This research has made use of the NASA/IPAC Extragalactic Database (NED)
which is operated by the Jet Propulsion Laboratory, California Institute
of Technology, under contract with the National Aeronautics and Space
Administration.

\begin{appendix}
\section{The significance of sources with a Gauss filter}

In this appendix we describe how we determine the significance of substructural
features in a Gauss filtered image. The process of Gauss filtering can be
described mathematically by:
\begin{eqnarray}
\lefteqn{G(x,y) =\left(\frac{1}{\sqrt{2\pi}\sigma}\right)^2} \nonumber \\
& &  \int_x \int_y e^{-(x-x^{\prime})^2/2\sigma^2} e^{-(y-y^{\prime})^2
/2\sigma^2}
f(x^{\prime},y^{\prime})\mbox{d}x^{\prime} \mbox{d}y^{\prime}
\end{eqnarray}
$x$ and $y$ are the coordinates of the filtered and $x^{\prime}$ and
$y^{\prime}$ the coordinates of the original image. The value of each pixel
is described by $f(x^{\prime},y^{\prime})$.
Applying the error propagation formalism to this function, assuming Poisson 
noise
(i.e. $\Delta f(x,y) = \sqrt{f(x,y)}$), we obtain:

\begin{eqnarray}
\lefteqn{ \Delta G(x,y) = \left(\frac{1}{\sqrt{2\pi}\sigma}\right)^2} \nonumber
\\
&  \left(\!
\int_x\!\!\int_y\!\! \left(\!e^{-(\!x\!-x^{\prime}\!)^2\!/\!2\!\sigma^2}\!\! 
e^{-(\!y\!-y^{\prime}\!)^2\!/\!2\!\sigma^2}\!\!
f^{1\!/\!2}(\!x^{\prime}\!,\!y^{\prime}\!)\!\right)\!^2\mbox{d}x^{\prime}\!
\mbox{d}y^{\prime}\right)\!\!^{1/2}
\end{eqnarray}

To realize this function one can apply a modified Gauss filter on the
original image with a filter size of
\begin{displaymath}
\sigma^{\prime} = \frac{\sigma}{\sqrt{2}}~~.
\end{displaymath}
and calculating the square root of this modified Gauss filtered image
for each pixel and multiplying it with a factor of
$\frac{1}{2\sqrt{\pi}\sigma}$. This  gives $\Delta G(x,y)$.
For the residual image the error $\Delta G(x,y)$ is the same as for the 
original smoothed image. Therefore 
to get the significance of the residuals one first calculates the subtracted
count rate image by $G(x,y)-$Modell$(x,y)$ and calculates the statistical error
of the residuals which is $\Delta G(x,y)$. The significance image is then
\begin{equation}
S(x,y) = \frac{G(x,y)-\mbox{Modell}(x,y)}{\
(x,y)}~~.
\end{equation}

\end{appendix}


\begin{thebibliography}{45}

\bibitem[]
{} Bardelli S., Zucca E., Malizia A., Zamorani G., Scaramella R.,
Vettolani, G., 1996, A\&A, 305, 435

\bibitem[]
{} Belloni P.,R\"oser H.-P., A\&AS, in press

\bibitem[]{}
Birkinshaw M., Gull S.F., Hardebeck H.E., 1984, Nature, 309, 34

\bibitem[]
{} Birkinshaw M., Hughes J.P., Arnaud K.A., 1991, ApJ, 379, 466

\bibitem[]{}
Briel U., Henry J.P., B\"ohringer H., 1992, A\&A, 259, L31

\bibitem[]
{} Buote D.A., Canizares C.R., 1992, ApJ, 400, 385

\bibitem[]
{} Butcher H., Oemler A., 1978, ApJ, 219, 18

\bibitem[]
{} Cavaliere A., Fusco-Femiano R., 1976, A\&A, 49, 137

\bibitem[]
{} Cavaliere A., Fusco-Femiano R., 1981, A\&A, 100, 194

\bibitem[]
{} Dickey J.M., Lockman F.J., 1990, ARA\&A, 219, 18

\bibitem[]
{} Dressler A., Gunn J.E. 1992, ApJS, 78, 1

\bibitem[]
{} Elbaz D., Arnaud M., B\"ohringer H., 1995, A\&A, 293, 337

\bibitem[]
{} Ellis R.S., Couch W.J., MacLaren I., Koo D.C., 1985, MNRAS, 217, 239

\bibitem[]
{} Evrard A.E., Metzler C.A., Navarro J.N., 1995, ApJ, submitted

\bibitem[]
{} Fabian A.C., Nulsen P.E.J., Canizares C.R., 1984 Nat, 310, 733

\bibitem[]
{} Fleming T.A., Giampapa M.S., Schmitt J.H.M.M., Bookbinder J.A., 1993, ApJ,
410, 387

\bibitem[]
{}Gerbal D., Durret F., Lima-Neto G., Lachi\'eze-Rey, 1992, A\&A, 253, 77

\bibitem[]
{}Gorenstein P.D., Fabricant D., Topka K., Harnden F.R., Tucker W.H., 1978,
ApJ, 224, 7188

\bibitem[]
{}Hughes J.P., Birkinshaw M., Huchra J.P., 1995, ApJ, 448, L93

\bibitem[]
{} Ikebe Y., 1995, Ph.D. Thesis, University of Tokyo

\bibitem[]
{}Jones C., Forman W., 1984, ApJ, 276, 38

\bibitem[]
{} Kaiser N., Squires G., 1993, ApJ, 404, 441

\bibitem[]
{} Koo D.C., 1981, ApJ, 251, 75

\bibitem[]
{} Luppino G.A., Kaiser N., 1996, ApJ, in press

\bibitem[]
{} Luppino G.A., Gioia I., 1995, ApJL, 445, L77

\bibitem[]
{} Mulchaey J.S., Davis D.S., Mushotzky R.F., Burstein D., 1996, ApJ, 456, 80

\bibitem[]
{}Neumann D.M., B\"ohringer H., 1995, A\&A, 301, 865

\bibitem[]
{} Ponman T.J., Bourner P.D.J., Ebeling H., B\"ohringer H., 1996, in: X-Ray 
Imaging and Spectroscopy of Cosmic Hot Plasmas, F. Makino (ed.) in press

\bibitem[]{}
Rephaeli Y., 1995, Ann. Rev. Astron. Astrophys., 33, 541

\bibitem[]{}
Rosati P., 1995, Ph.D. thesis, University of Rome

\bibitem[]
{} Sarazin C.L., Bahcall J.N., 1977, ApJS 34, 451

\bibitem[]
{} Schindler S., M\"uller E., 1993, A\&A, 272, 137

\bibitem[]
{} Schindler S., 1995, A\&A, 305, 756

\bibitem[]
{} Smail I., Ellis R.S., Fitchett M.J. Edge A.C., 1995, MNRAS, 273, 277

\bibitem[]
{} Squires G., Neumann D. M., Kaiser N., Arnaud M., Babul A., B\"ohringer
H., Fahlman G., Woods D., 1996 submitted

\bibitem[]
{} Tr\"umper J., 1984, Adv. Space Res., 2, 241

\bibitem[]
{} Tr\"umper J., 1992, QJRAS, 33, 165

\bibitem[]
{} Tsuru T., Koyama K., Hughes J.P., Arimoto N., Kii T., Hattori M., 1996,
in: UV and X-ray Spectroscopy of Astrophysical and Labatory Plasmas, Yamashita
K., Watanabe T. (eds.), Universal Academic Press, Tokyo, p.375

\bibitem[]{}
Uson, J.M., 1986, Proc. of the conference on Radio Continuum Processes
in Clusters of Galaxies, Green Bank, NRAO, p. 255

\bibitem[]
{} Van Haarlem M., Van de Weygaert R., 1993, ApJ, 418, 544

\bibitem[]
{} Wirth G.D., Koo D.C., Kron R.G., 1994, ApJ, 435, L105

\bibitem[]
{} White S.D.M., Navarro J.F., Evrard A.E., Frenk C.S., 1993, Nat,
366, 429

\bibitem[]
{} Yamashita K., 1994, in {\it CLUSTERS OF GALAXIES} eds. Durret, F., Mazure,
A., Tr\^{a}n  Than V\^{a}n, J., EDITIONS FRONTIERES, Gif-sur-Yvette, p.153

\bibitem[]
{} Zimmermann H.U., Becker W., Belloni T., D\"obereiner S., Izzo C., Kahabka
P., Schwendtker O., 1994, EXSAS User's Guide, MPE Report 257

\end{thebibliography}
\end{document}